\documentclass[useAMS,usenatbib]{mnras}
\usepackage{graphicx,psfig,cite,epsf}  
\usepackage{times}
\usepackage{subfigure}
\usepackage{textcomp}
\usepackage{multirow}
\usepackage{amsmath}	
\usepackage{amssymb}	
\usepackage{gensymb}

\def\apjl{ApJ}                
\def\apjs{ApJS}               
\def\ao{Appl.~Opt.}           
\def\aap{A\&A}                

\def\swift{{\it Swift/XRT}}
\def\chandra{{\it Chandra}}
\def\xmm{{\it XMM-Newton}}
\def\hst{{\it HST}}

\title[A new ULX in NGC 5907]{A new Ultraluminous X-ray source in the galaxy NGC 5907}

\author[F. Pintore et al.]{F. Pintore$^1$\thanks{E-mail: pintore@iasf-milano.inaf.it}, A. Belfiore$^1$, G. Novara$^{1,2}$, R. Salvaterra$^1$, M. Marelli$^1$, A. De Luca$^1$, \newauthor  M. Rigoselli$^{1,3}$, G. Israel$^4$, G. Rodriguez$^4$, S. Mereghetti$^1$, A. Wolter$^5$, D. J. Walton$^6$, \newauthor F. Fuerst$^7$,  E. Ambrosi$^{8,9}$, L. Zampieri$^8$,  A. Tiengo$^{1,2}$,   C. Salvaggio$^{3}$ \\
$^1$ INAF -- IASF Milano, Via E. Bassini 15, 20133 Milano, Italy; \\
$^2$ Scuola Universitaria Superiore IUSS Pavia, Piazza della Vittoria 15, 27100 Pavia, Italy; \\
$^3$ Dipartimento di Fisica, Universit\`a degli Studi di Milano-Bicocca, Piazza della Scienza 3, I-20126 Milano, Italy; \\
$^4$ INAF - Osservatorio astronomico di Roma, Via Frascati 44, I-00040, Monteporzio Catone, Italy; \\ 
$^5$ INAF, Osservatorio Astronomico di Brera, via Brera 28, 20121 Milano, Italy; \\
$^6$ Institute of Astronomy, Madingley Road, Cambridge CB3 0HA, UK \\
$^7$  European Space Astronomy Centre (ESAC), Science Operations Departement, 28692 Villanueva de la Cañada, Madrid, Spain; \\
$^8$ INAF-Osservatorio Astronomico di Padova, Vicolo dell'Osservatorio 5, I-35122 Padova, Italy \\
$^9$ Dipartimento di Fisica e Astronomia, Universit\`a degli Studi di Padova, Via VIII Febbraio 1848, 2, 35122 Padova, Italy }

\date{Accepted  . Received  ;}

\pubyear{2017}

\begin{document}

\maketitle

\begin{abstract}
We report on the serendipitous discovery of a new transient in NGC 5907, at a peak luminosity of $6.4\times10^{39}$ erg s$^{-1}$. The source was undetected in previous 2012 \chandra\ observations with a $3\sigma$ upper limit on the luminosity of $1.5\times10^{38}$ erg s$^{-1}$, implying a flux increase of a factor of $>35$. 
We analyzed three recent 60ks/50ks \chandra\ and 50ks \xmm\ observations, as well as all the available \swift\ observations performed between August 2017/March 2018. Until the first half of October 2017, \swift\ observations do not show any emission from the source. 
The transient entered the ULX regime in less than two weeks and its outburst was still on-going at the end of February 2018. The 0.3--10 keV spectrum is consistent with a single multicolour blackbody disc (kT$\sim1.5$ keV). 
The source might be a $\sim$30 M$_{\odot}$ black hole accreting at the Eddington limit. However, although we did not find evidence of pulsations, we cannot rule-out the possibility that this ULX hosts an accreting neutron star.

\end{abstract}

\section{Introduction}

Ultraluminous X-ray sources (ULXs, e.g. \citealt{kaaret17}) are a class of extragalactic point like sources, with X-ray luminosities higher than $10^{39}$ erg s$^{-1}$ up to, in the most extreme cases, $10^{42}$ erg s$^{-1}$ (e.g. HLX-1; \citealt{farrell09}). Such luminosities are hence well above the Eddington limit for accretion of pure hydrogen onto a 10 M$_{\odot}$ black hole (BH). The number of known ULXs is still increasing and nowadays more than 300 sources are confirmed ULXs, as catalogued e.g. by ROSAT \citep{liu05}, \chandra\ \citep{swartz11} and {\it XMM-Newton} \citep{walton11}.  

ULXs represent a current hot topic of X-ray astronomy because of their peculiar accretion properties. It is believed that ULXs are accreting compact objects in binary systems \citep[e.g.][]{motch14,urquhart18}, but the nature of the compact object is still under debate: indeed ULXs may host either i) intermediate mass BH (IMBHs, $100-10^5$ M$_{\sun}$; \citealt{colbert99}) accreting well below the Eddington limit, or ii) stellar BHs (5 M$_{\sun}<$ M$_{BH} < 80$ M$_{\sun}$; e.g. \citealt{zampieri09, gladstone09, walton13,middleton15}) or iii) even neutron stars (NSs), accreting at extremely super Eddington rates \citep[e.g.][]{israel16a}. 

\begin{figure*}
\center
		\includegraphics[width=15.0cm]{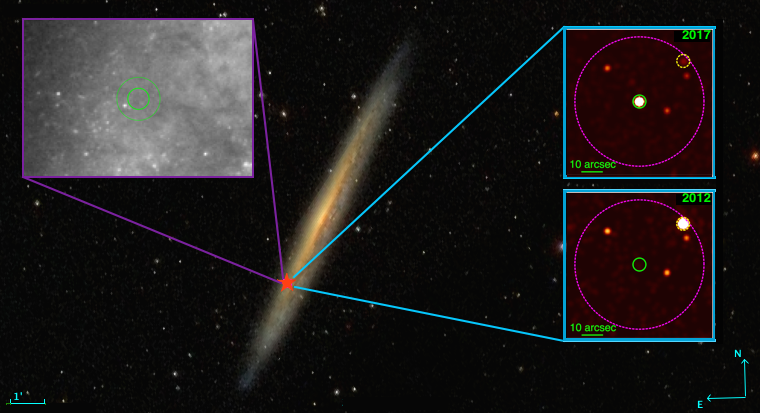}
   \caption{Main figure: SDSS optical image of the galaxy NGC 5907 (north is up); the red star indicates the {\it Chandra} position of the new ULX transient. Inset box-right: \chandra\ images of the 2012 (bottom) and 2017 (top) observations, where the green, dashed purple and dashed yellow circles indicate the \chandra\ (3'' radius) and EPIC (30'' radius) extraction regions for the new transient, and the ULX-1 position, respectively. 
   Inset box-left: Archival \hst/WFPC2 image ($15\arcsec \times10\arcsec$) of the field of the new transient source NGC 5907 ULX-2  (F606W filter); ULX-1 is out of image. The green circles mark the X-ray uncertainty region (radius of $0.47\arcsec$ and $1.42\arcsec$ at 68\% and 99\% confidence level, respectively) after registering Chandra astrometry using a set of SDSS sources. The brightest object within the 99\% region (F606W magnitude of $24.65\pm0.07$, ST system) is the candidate optical counterpart of the transient ULX. }  
   \label{image}
\end{figure*}

However, it is now thought that the vast majority of the ULX population is  composed of super-Eddington accreting compact objects. In particular, the presence of NSs in at least 4 ULXs is confirmed by the detection of pulsation (\citealt{bachetti14,israel16a}a; \citealt{fuerst16a,israel16b}b, \citealt{carpano18a}). In addition, these pulsating ULXs (PULXs) show significant flux variability of several orders of magnitude.
The variety of possible compact objects in ULXs suggests they are a manifold population hosting both BHs and NSs \citep[e.g.][]{middleton17}, although their relative number is still highly uncertain. 

Here we report the discovery of a new X-ray transient in the direction of the galaxy NGC 5907 (Figure~\ref{image}), located at a distance of 17.1 Mpc \citep{tully13}. NGC 5907 hosts a large population of X-ray sources, with the brightest being the PULX NGC 5907 X-1 \citep[e.g.][]{sutton13b,walton17,israel16a}. The new source reported here could be the second ULX hosted in this galaxy. 

\section{data reduction}
\label{data_reduction}

\noindent {\bf Chandra.} {We analyzed two \chandra\ ACIS-S archival observations of 11th February 2012 (Obs.ID: 12987, 14391). We also obtained, as Director Discretionary Time (DDT), three ACIS-S observations (Obs.ID: 20830, 20994, 20995) taken on 7th November 2017, 28th February and 1st March 2018. The Chandra observation exposure times are $\sim17$ ks, $\sim14$ ks, $\sim52$ ks, $\sim32$ ks and $\sim16$ ks, respectively.} \chandra\ data were reduced with {\sc ciao} v.4.9 and calibration files CALDB v.4.7.6.  
The source events were chosen from a circular region of radius 3'', while the background was selected in a circular region of radius 15'' and free of sources.
We obtained source spectra with the {\sc ciao} task SPECEXTRACT, which creates the appropriate response/auxiliary files for the spectral analysis. With the {\sc pileup\_map} tool, we estimated that pile-up effects contribute to $\sim$1$\%$, and we can neglect it. {Because of their close temporal proximity, the 2018 observations were analyzed as a single dataset.}

\vspace{0.1cm}
\noindent {\bf Swift.} We analyzed all the available (24) XRT observations of the {\it Neil Gehrels Swift Observatory} taken between August 2017 and {March 2018}, with an average exposure time of $\sim2$ ks. These observations are generally spaced at intervals of about one week. We reduced the data with {\sc xrtpipeline} and extracted 0.3--10 keV lightcurves selecting circular regions of radii 20'' and 130'' for source and background, respectively. We also analyzed UVOT data by reducing them with standard procedures\footnote{\url {http://www.swift.ac.uk/analysis/uvot/index.php}}.

\vspace{0.1cm}
\noindent {\bf XMM-Newton.} On 2nd December 2017, we obtained a DDT \xmm\ observation with a total exposure time of 60 ks. We extracted the data from EPIC-pn and the two EPIC-MOS cameras, and reduced them with SAS v15.0.0, We selected single- and double-pixel events ({\sc pattern}$\leq$4), and single- and multiple-pixel events ({\sc pattern}$\leq$12), for pn and the MOS, respectively. We removed high background time intervals and we obtained net exposure times of $\sim$38 ks in the pn and $\sim$58 ks in the MOS. For spectral and timing analyses, we extracted the source and background events from circular regions of radii 30'' and 60'', respectively. 

\vspace{0.1cm}
\noindent {\bf HST.} 
We analysed optical images of the field collected with the {\em Hubble} Space Telescope (HST). The field was observed  with the Wide Field and Planetary Camera 2 (WFPC2) instrument on 1996, March 31 in the F450W ($\lambda=4556\AA$, $\Delta \lambda=951\AA$) and F814W ($\lambda=8012\AA$, $\Delta \lambda=1539\AA$) filters, and on 2007, October 1 in the F606W filter ($\lambda=5997\AA$, $\Delta \lambda=1502\AA$), with exposure times of 2340 s, 960 s and 3400 s, respectively. We retrieved calibrated, geometrically-corrected images from the {\em Hubble} Legacy Archive (HLA\footnote{\url {https://hla.stsci.edu/hlaview.html}}) -- such images have an improved astrometry based on cross-correlation with the {\em Sloan} Digital Sky Survey catalogue, with r.m.s. accuracy of $\sim0.1"$ per coordinate. We carried out a source detection using the SExtractor software \citep{bertin96} and we converted count rates to magnitudes in the ST system using the photometric calibration provided by the HLA pipeline.

\begin{figure*}
\center
\hspace{-0.5cm}
\includegraphics[width=7.2cm]{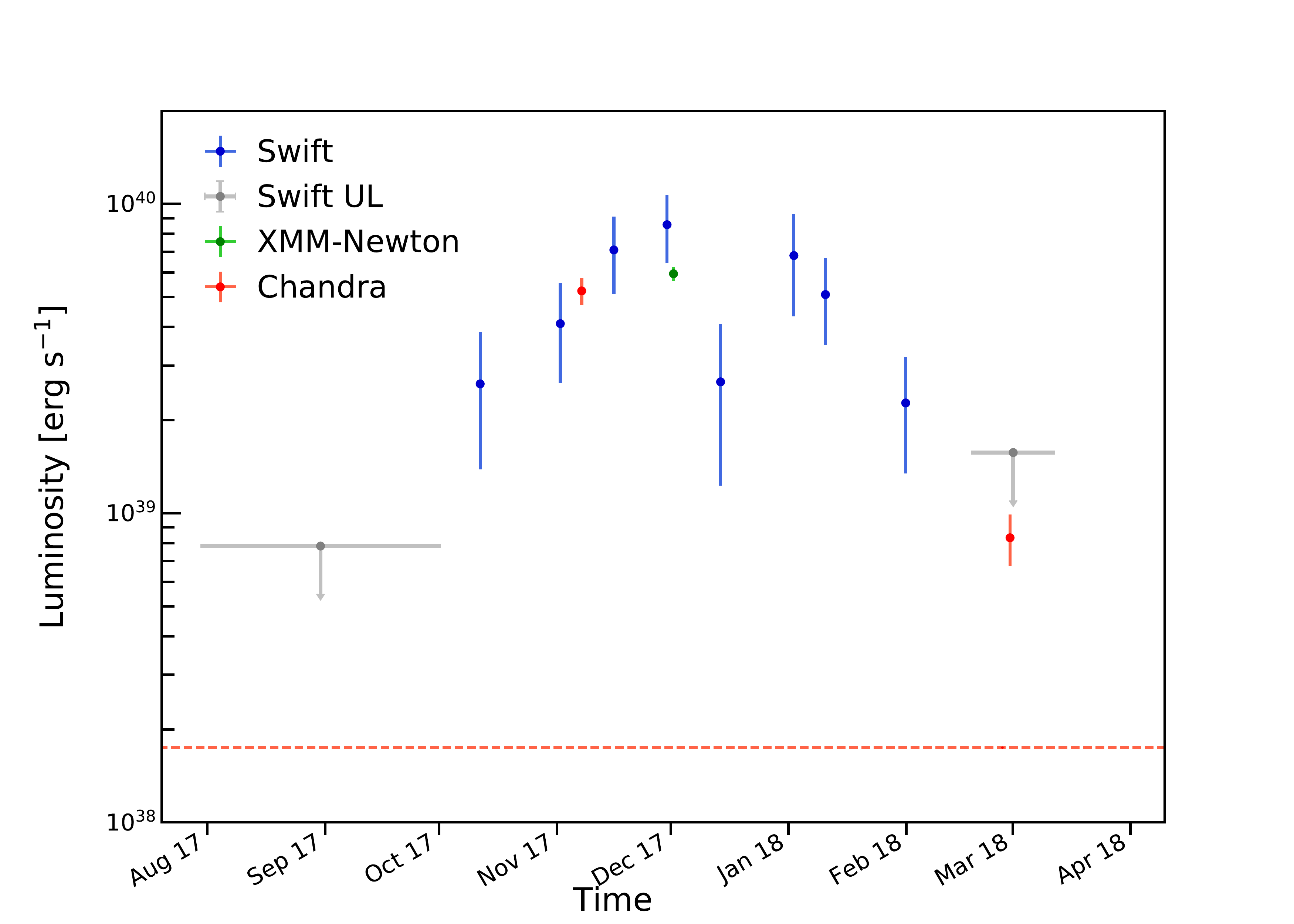}
\hspace{-0.5cm}
\includegraphics[width=4.2cm,height=4.6cm]{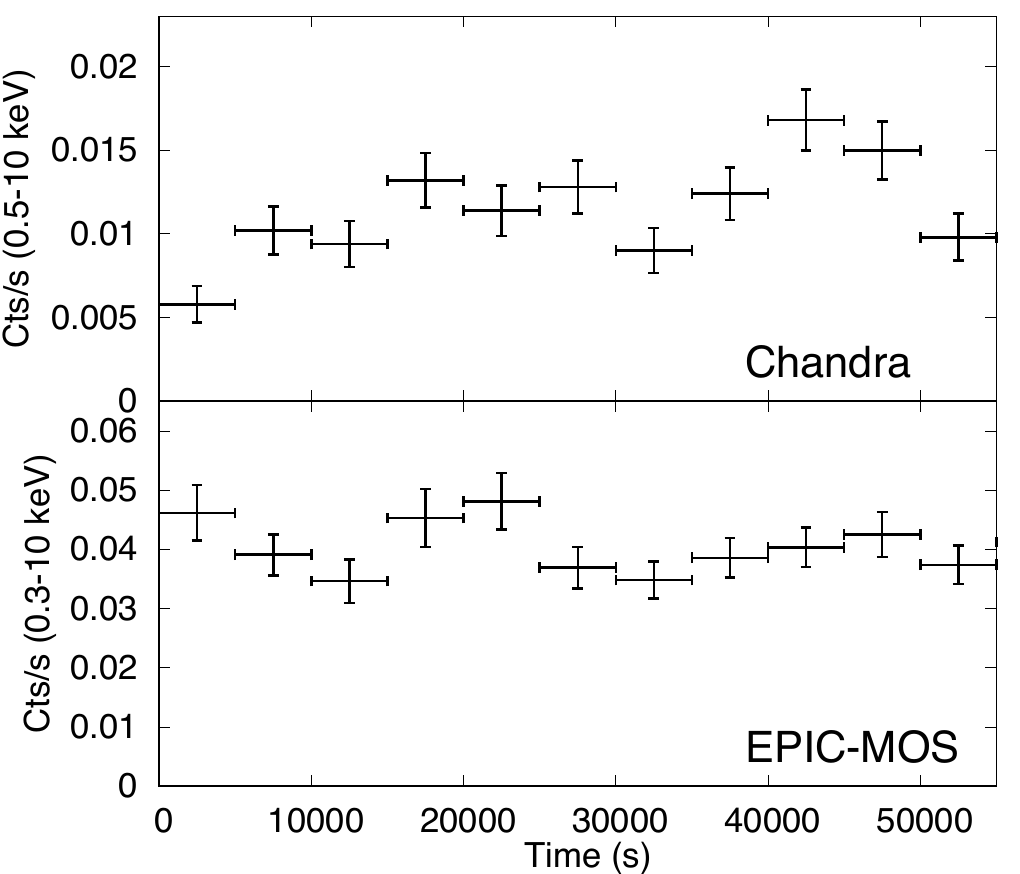}
\hspace{0.1cm}
\includegraphics[width=6.3cm]{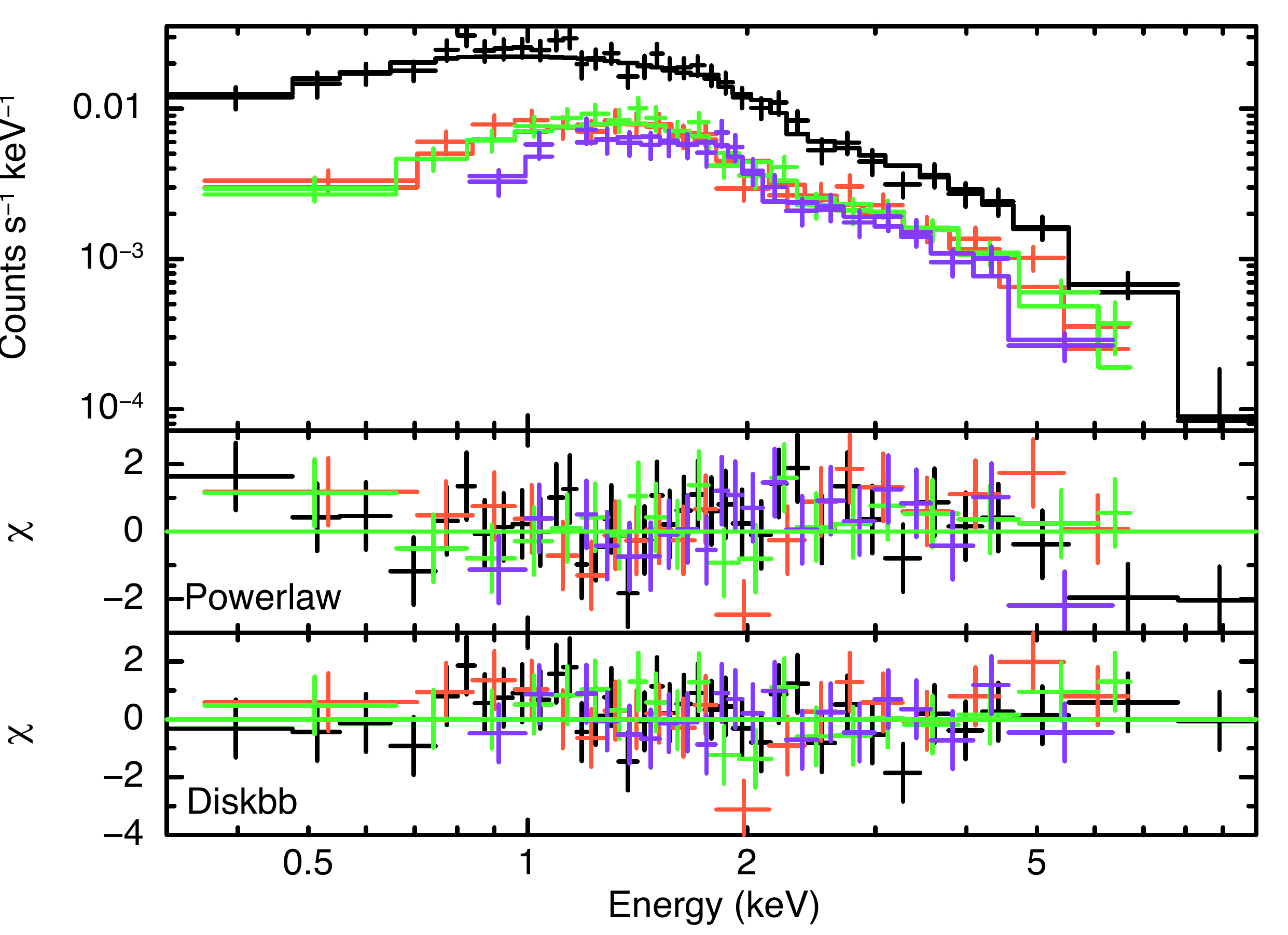}
   \caption{{\it \textbf{Left}}: background-subtracted 0.3--10 keV unabsorbed luminosity lightcurve of ULX-2; the blue, green and red points represent the \swift\ (binning two consecutive observations in each point; errors at 1$\sigma$), \chandra\ and \xmm\ observations (errors at $90\%$); the grey points are instead the $3\sigma$ \swift\ upper limits (where we assume the EPIC {\sc diskbb} best fit model) obtained stacking all the observations between August--September 2017 and February--March 2018; finally, the dashed red line is the 2012 \chandra\ upper limit. {\it \textbf{Center}}: 2017 \chandra\ (top) and EPIC/MOS (bottom) background subtracted lightcurve, binned with 5000s (to have more than 20 counts in each bin), with the X-axis reported to the starting time of each observation. {\it \textbf{Right}}: top panel: averaged spectrum of the \xmm\ (black: EPIC-pn, red: EPIC-MOS1, green: EPIC-MOS2) and 2017 \chandra\  (blue) observations, fitted with a {\sc diskbb}; middle panel: residuals of the {\sc powerlaw} fit; bottom panel: residuals of the {\sc diskbb} fit; spectra have been rebinned for display purposes only.}  
        \label{ch_lc}
\end{figure*}

\section{Results}

In the 2017 \chandra\ ACIS-S image, we detected a new bright X-ray source which was not present in the same field in the 2012 \chandra\ observation (right inset in Figure~\ref{image}). The most precise position of this new transient (ULX-2 hereafter) was obtained with the {\sc ciao} task WAVDETECT, which gives coordinates RA = 15h16m1.10s and Dec = 56\textdegree17'51.40'' (statistical uncertainty of 0.03''). The source is therefore superimposed on the NGC 5907 galactic plane and at a distance from ULX-1  of $\sim28$''. 

\subsection{Temporal analysis}

In the {2012 \chandra\ } observations, ULX-2 was not detected, with a $3\sigma$ upper limit of $4.4\times10^{-15}$ erg cm$^{-2}$ s$^{-1}$ on the 0.3--10 keV flux (assuming a multicolour blackbody disc, see below).
The transient was undetected also in all the \swift\ observations taken before 15th October 2017. Stacking the August/September observations we obtained a 3$\sigma$ upper limit of $2.2\times10^{-14}$ erg cm$^{-2}$ s$^{-1}$ on the average flux (shown in Figure~\ref{ch_lc}-left). A source consistent with the position of ULX-2 was detected in later \swift\ observations taken at the end of October till {the half of February} 2018. Because of the angular resolution of \swift\ ($\sim$18'' HPD), we cannot exclude that other sources, very close to ULX-2, became active in the same period. However, if we associate the \swift\ source with ULX-2, these findings suggest that it entered the ULX regime in the second half of October 2017. 

The 0.3--10 keV background-subtracted 2017 \chandra\ lightcurve of ULX-2, is shown in Figure~\ref{ch_lc}-right (top panel).  A fit with a constant flux gives $\chi^2/dof=46/10$, indicating the presence of variability (for timescales $>$5000s). On the other hand, we found that, both during the {\it XMM-Newton} and {2018} {\it Chandra} observations, the source had a rather constant flux {($\chi^2/dof=12.1/11$ and $\chi^2/dof=5.3/6$;} Fig.~\ref{ch_lc}-center, bottom panel). 
We also investigated, in both XMM and \chandra\ data, the hardness ratio between the energy bands 0.5--1.5 keV ({\it soft}) and 1.5--10 keV ({\it hard}) which indicates no significant spectral variability. 

We then created a power spectrum (only for the \xmm\ and {2017} \chandra\ observations, because of the too low 2018 \chandra\ counting statistics) for the whole energy band (0.3--10 keV) but we could not find any coherent signal in both datasets. Assuming sinusoidal modulation, we derived a $3\sigma$ upper limit on the pulsed fraction (defined as the semi-amplitude of the sinusoid divided by the source average count rate) of 60$\%$ and 35$\%$ for \chandra\ and \xmm, for periods in the range $\sim$6s-5000s and $\sim$0.140s-150s, respectively. An accelerated search for coherent signals using a grid of 6234 $\dot{\text{P}}$/P values in the range $\pm{10^{-11}-10^{-6}}$ s s$^{-2}$ gave no statistically significant signals.

\subsection{Spectral analysis}

As the hardness ratio does not indicate significant spectral variability in the single observations, we extracted an averaged spectrum for the \xmm/EPIC (pn + MOS1/2), and the {2017 and 2018 \chandra\ data; they collect $\sim$$3600$, and $\sim$$550$ and $\sim$95 net source counts}, respectively. We rebinned the spectra with at least 20 counts per bin and used {\sc xspec} v.12.8.2 to fit them. 

We first analyzed the EPIC spectra and adopted simple models as a {\sc powerlaw}, or a {\sc blackbody} or a {\sc diskbb}, absorbed with {\sc tbabs} and using the abundances of \citet{wilms00}. We found that acceptable spectral fits ($\chi^2_{\nu} < 1$) are obtained with the {\sc powerlaw} or the {\sc diskbb} (Table~\ref{table_spectra}). However, the {\sc powerlaw} model left some skewed residuals (in particular above 5 keV, see Figure~\ref{ch_lc}, right-middle panel). Hence, in the following, we focus on the {\sc diskbb} fit (see Figure~\ref{ch_lc}, right-bottom panel). With the latter, we estimated a mean disc temperature of $\sim1.5$ keV, a total column density of $8\times10^{20}$ cm$^{-2}$, hence higher than the Galactic n$_{\text{H}}$ along the source direction ($1.38\times10^{20}$ cm$^{-2}$; \citealt{dickey90}), and an absorbed 0.3--10 keV flux of $(1.7\pm0.1) \times10^{-13}$ erg cm$^{-2}$ s$^{-1}$. If ULX-2 is in NGC 5907, the flux translates to an unabsorbed luminosity of $\sim$$6.4\times10^{39}$ erg s$^{-1}$ (for a distance of 17.1 Mpc).

\begin{table*}
\center
          \caption{Best-fit spectral parameters of the transient source. Errors at $90\%$ confidence level for each parameter of interest.} 
\scalebox{0.88}{\begin{minipage}{24cm}
      \label{table_spectra}
\begin{tabular}{lccccccccc}
\hline
&   & \multicolumn{3}{c}{XMM-Newton}  & \multicolumn{2}{c}{2017 \chandra} & \multicolumn{2}{c}{2018 \chandra} & \\
&  nH  & kT / $\Gamma$ & Norm. & Flux$^a$ & kT / $\Gamma$ & Norm. & kT / $\Gamma$ & Norm.  & $\chi^2/dof$$^b$\\
&  $10^{22}$ cm$^{-2}$ & (keV) &  & $10^{-13}$ erg cm$^{-2}$ s$^{-1}$ & (keV) &  & (keV) &   & \\
\hline
{\sc bbody} & $<0.8$ &$0.67^{+0.03}_{-0.03}$ & $1.6^{+0.1}_{-0.1} \times 10^{-6}$ & $1.36^{+0.08}_{-0.08}$ & $0.71^{+0.05}_{-0.05}$ & $1.3^{+0.1}_{-0.1} \times 10^{-6}$ & $0.6^{+0.1}_{-0.1}$ & $2.3^{+0.5}_{-0.4} \times 10^{-7}$  & [309/170]\\
{\sc powerlaw} & $0.28^{+0.05}_{-0.04}$ & $1.75^{+0.09}_{-0.1}$ & $3.9^{+0.4}_{-0.4} \times 10^{-5}$ & $1.9^{+0.1}_{-0.1}$ & $1.9^{+0.1}_{-0.1}$ & $3.2^{+0.4}_{-0.4} \times 10^{-5}$ & $1.9^{+0.4}_{-0.3}$ & $5.6^{+1.7}_{-1.5} \times 10^{-6}$  & {\bf 163/170} \\
{\sc diskbb} &  $0.08^{+0.03}_{-0.02}$ & $1.5^{+0.1}_{-0.1}$ & $1.7^{+0.5}_{-0.4} \times 10^{-3}$ & $1.7^{+0.1}_{-0.1}$ & $1.3^{+0.2}_{-0.1}$ & $2.2^{+1.1}_{-0.8} \times 10^{-3}$ & $1.1^{+0.3}_{-0.2}$ & $9.1^{+15}_{-5.9} \times 10^{-4}$  & {\bf 162/170} \\
\hline

\end{tabular}
\end{minipage}}
\flushleft $^a$ EPIC absorbed flux in the 0.3--10 keV energy band; $^b$ As found in EPIC; values in boldface are statistically acceptable.
\end{table*}

The {2017} \chandra\ spectrum was then fitted with the {\sc diskbb} model with n$_{\text{H}}$ fixed to the column density found in the EPIC spectra. The {\sc diskbb} provides a very good fit ($\chi^2/dof<1$), with a disc temperature consistent within the uncertainties with the EPIC value (Table~\ref{table_spectra}).
We note that the average ULX-2 absorbed flux during this \chandra\ observation was $\sim20\%$ lower ($(1.2\pm0.1) \times10^{-13}$ erg cm$^{-2}$ s$^{-1}$) than that found during the \xmm\ observation. However, we also checked that the source flux in the last 20 ks of this \chandra\ observation was  consistent with the \xmm\ one. 
We also note that, inside the \xmm\ source extraction region, at least three other weaker sources are included (as seen in the \chandra\ image). We estimated that they may increase the ULX-2 flux of $\leq$$10\%$ and can possibly affect its spectrum but the contaminant effect on the ULX-2 EPIC spectrum appears very marginal because of the strong agreement with the \chandra\ data.

{Finally, the 2018 \chandra\ spectrum was still compatible with a {\sc diskbb} model ($\chi^2_{\nu}/dof=5.33/6$, Table~\ref{table_spectra}). Fixing the column density to $8\times10^{20}$ cm$^{-2}$, we found an inner disc temperature of $1.1\pm0.3$ keV and an absorbed 0.3--10 keV flux of $(2.1\pm0.4) \times10^{-14}$ erg cm$^{-2}$ s$^{-1}$, i.e. a luminosity of $\sim7\times10^{38}$ erg s$^{-1}$ and thus well below the ULX threshold luminosity.}

\subsection{Optical data}
We searched for an optical counterpart inside the X-ray \chandra\ error circle (radius of 1.3''). In the archival HST/WFPC2 data observations, we found a {bright} object (Figure~\ref{image}, inset-right) that we consider as the most promising candidate counterpart, with ST magnitudes of $23.7\pm0.1$, $24.65\pm0.07$ and $24.12\pm0.09$ {(absolute magnitudes of $-7.5\pm0.1$, $-6.51\pm0.07$ and $-7.04\pm0.09$)} in the F450W, F606W and F814W filters, respectively. Luminosity and colours of this source are consistent with an OB-type star in NGC 5907. We compared the simultaneous F450W and F814W measurements with emission models of ULX binaries (Ambrosi \& Zampieri, in prep.). The evolutionary tracks of a binary with a $\sim$$20-30 M_\odot$ BH and an evolved donor with a main sequence mass of $\sim$$20 M_\odot$ are in agreement with the Galactic extinction corrected photometry, and the mass transfer rate is above Eddington. A detailed modelling is under way and will be presented elsewhere.

\section{Discussion}
The long-term variability, the n$_\text{H}$ higher than the Galactic column density and the quite hard X-ray spectrum of the newly discovered source disfavour the identification with a star in our Galaxy. In particular, a foreground star of any spectral type and with the X-ray flux of ULX-2 should have an optical magnitude M$_V < 18$ mag \citep{krautter99}. To explore the chance coincidence of a background AGN, we considered the X-ray Log N -- Log S of extragalactic sources: the number of AGN per square degree with a flux larger than $>10^{-13}$ erg cm$^{-2}$ s$^{-1}$ is of the order of a few \citep[e.g.][]{moretti03}. For an estimated area of  4.7'$\times$0.5' covered by NGC 5907, this translates on a number of contaminant AGN less than $5\times10^{-3}$. We finally rule-out the possibility of a supernova event as no significant UV variability has been seen in UVOT data, where, however, the source contribution cannot be distinguished from the rest of the galaxy.

We therefore conclude that the new source is indeed in all respects a ULX transient in NGC 5907, and its peak luminosity is $\sim6.4\times10^{39}$ erg s$^{-1}$, making this source the second brightest ULX in the galaxy. This implies a luminosity increment of at least a factor of $\sim$35 with respect to the 2012 \chandra\ upper limit. {Before October 2017, it was below the nominal ULX threshold luminosity ($>10^{39}$ erg s$^{-1}$) where it returned at the end of February 2018 (L$_X=7\times10^{38}$ erg s$^{-1}$), following an exponential decay with folding time of $27\pm4$ days}.   
We remark that we could investigate the new source properties with \xmm\ because the close ($\sim28$'') ULX-1 was in quiescence and hence not contaminating the ULX-2 emission. 
Although beyond the scope of this paper, we double-checked all the available \xmm\ observations of ULX-1 and, through a Maximum Likelihood analysis based on the EPIC pn/MOS PSFs \citep[e.g.][]{rigoselli18}
we found hints for the presence of ULX-2 only during the observation of 6 November 2013 (at estimated 0.3--10 keV flux of $\sim$$8\times10^{-15}$ erg cm$^{-2}$ s$^{-1}$), when ULX-1 was very weak. However, our analysis does not fully account for possible contaminants close to the ULX-2 position, that may affect the reported flux.
 
ULX spectra, in the so called Ultraluminous state \citep{roberts07}, are generally well described by at least two spectral components interpreted as a soft thermal emission (kT$\sim0.1-0.5$ keV) from cold optically thick outflows and hard thermal emission from a hot modified accretion disc or corona close to the compact object \citep[e.g.][]{gladstone09,pintore12,sutton13,bachetti14,middleton15,walton17}. 
{On the other hand, there is a small sample of ULXs with very limited data quality displaying properties consistent with the spectral states of Galactic accreting black holes X-ray binaries \citep{sutton12}. However, some of these ULXs showed the features of the Ultraluminous state when data of better quality became available \citep[e.g.][]{sutton13,pintore16}. We believe that this is not the case for ULX-2 because the quality of its current spectra is relatively good ($\sim$3500 net EPIC counts).}

We found that the average ULX-2 spectrum is described by a single multicolour accretion disc with a temperature of $\sim$1.5 keV, indicating that the source may be in a {\it soft} state reminiscent of those found in Galactic accreting BH binaries \citep[e.g.][]{mcclintock06}. However, although not supported by the data, we cannot exclude that its disk-like shape might suggest that ULX-2 is in the super-Eddington ``very thick-state'' {(i.e. a spectral shape consistent with Comtponization from an optically thick corona with $\tau$$>$10 and kT$_e$$\sim$1--2 keV}; see \citealt{pintore12}). In any case, if the source is radiating close to the Eddington limit (and we consider the {\it soft} state as reliable), the implied mass would be M$_{BH} \sim L_{\text{X}} / (1.38\times10^{38}) \text{M}_{\odot} \sim 30$ M$_{\odot}$ (assuming isotropic emission). Although the mass of such an object would be larger than any accreting BH found in our Galaxy, the LIGO and VIRGO detectors showed the existence of BHs of similar sizes \citep{abbott16,abbott17}. Assuming an inclination angle $<$70\textdegree\ (because of the absence of dips or eclipses), the best-fit {\sc diskbb} normalization corresponds to an estimated inner accretion disc radius of $<$140 km, or to an upper limit of $\sim$15 M$_{\odot}$ for the mass of a non-rotating BH \citep{lz09}. {In addition, although the error bars are large, we note that the relation  L$_X\propto$ kT$_{disc}^4$ (valid for a standard accretion disc) is compatible with our temperature and luminosity values.}

Until now, only a handful of transient ULXs are known and not all of them have been deeply studied. Some examples of ULX transients are M31 ULX-1 and ULX-2 \citep[e.g.][]{middleton12,esposito13,middleton13}, three ULXs in the Cartwheel galaxy \citep{crivellari09}, CXOU 133705.1-295207 in M83 \citep{soria12}, CXOU J132518.2-430304 in NGC 5128 \citep{sivakoff08} and M82 X-1. {Their outbursts can be explained, for instance, in terms of an unstable mass transfer from the companion star.} The three PULXs -- M82 X-2 (e.g. \citealt{bachetti13}), NGC 7793 X-1 \citep{motch14} and NGC 5907 ULX-1 \citep[e.g.][]{walton15} -- from time to time also show high flux variability of several orders of magnitude, {which may be explained by the onset/switch-off of the propeller stage.} {The flux variability of at least a factor of $\sim$40 between high and low flux states of ULX-2 is similar to that ascribed to the propeller effects in M82 X-2 \citep[see e.g.][]{tsygankov16}.} 
We did not find any coherent pulsations in ULX-2 with an upper limit on the pulsed fraction of $35\%$. Only the pulsed fraction of NGC 300 ULX-1 is larger than this value, while the other three PULXs have smaller modulations.
Thus, we cannot exclude that ULX-2 hosts a NS accreting at very high rates although, from a spectral point of view, it may be slightly softer than the PULXs \citep[e.g.][]{pintore17,walton18}. Furthermore, all the PULXs show super-orbital modulations that the available data do not allow us to search. But a deeper monitoring with {\it Swift} or \xmm\ (during the quiescence of ULX-1) or, even better, with \chandra\ could provide indications about super-orbital modulations. 

The \hst\ image indicates that the source might have an optical counterpart with the colours of an OB type star. Only $\sim$20 ULX optical counterparts have been identified \citep{tao11,gladstone13} and, when individual optical counterparts can be associated with a ULX, they often appear to have the luminosities and colours of OB type giants or supergiants \citep[e.g.][]{zampieri04,soria05}, even though the disc emission may contribute significantly to the optical flux \citep{patruno08}. New optical observations are strongly needed to constrain the nature of the ULX-2 counterpart. 

\section*{Acknowledgements} 
We thank Dr. B. Wilkes, Dr. N. Schartel and Dr. Brad Chenko who made the {\it Chandra}, {\it XMM-Newton}  and the {\it Neil Gehrels Swift Observatory} DDT observations possible. 

\noindent This work is based on NASA/ESA Hubble Space Telescope observations and obtained from the Hubble Legacy Archive. 

\noindent DJW acknowledges support from STFC in the form of an Ernest Rutherford fellowship. 
We acknowledge financial support from the Italian National Institute for Astrophysics (INAF) through the project ``ACDC - ASTRI/CTA Data Challenge'', from the Italian Space Agency (ASI) through the ASI-INAF agreements 2015-023-R.0 and 2017-14-H.0, as well as from the EXTraS project (``Exploring the X-ray Transient and variable Sky''), funded from the EU's 17$^{th}$ Framework Programme under grant agreement no. 607452, and the high performance computing resources and support provided by the CINECA (MARCONI) and by the INAF - CHIPP

\addcontentsline{toc}{section}{Bibliography}
\bibliographystyle{mn2e}
\bibliography{biblio}

\bsp
\label{lastpage}
\end{document}